\newcommand\eq[1]{Eq.~(\ref{#1})}

\documentclass[12pt]{article}
\usepackage{amssymb,graphics}
\begin{document}

\title{Comment on ``Quantum noise influencing human behaviour could fake
 effectiveness of drugs in clinical trials''}
\author{{\em Alexander Yu.\ Vlasov}\\
{\small FRC/IRH, 197101 Mira Street 8, St.--Petersburg, Russia}}
\date{}
\maketitle
\begin{abstract}
 Here are discussed some problems concerning quant-ph/0208006.
\end{abstract}

\section{Introduction}

In the paper \cite{clinic} it is suggested to consider possibility of
influation of quantum effects in every-day life events like clinical
studies. If to recall, how difficult and fine may be experiments for
testing of quantum nonlocality \cite{Aspect}, it is really unexpectedly, how
it would be possible to test such kind of results in very complicated
area, like clinical studies of real people, there even
classical statistical criteria may be applied only with certain limitations.
It should be mentioned, that even in precise physical experiments
observation of quantum correlations still is under certain investigation
\cite{GisLoop1,GisLoop2}.
Let us consider a few concrete problems related with models discussed
in \cite{clinic}.

\section{Physics and Cryptography. (Testing quantum correlation for
time-like interval)}

It should be mentioned, that first design used in \cite{clinic} has essential
difference with usual experiment \cite{Aspect,GisLoop1} and to talk about
experimental confirmation is not quite correct.

\begin{figure}[ht]
\begin{center}
\includegraphics{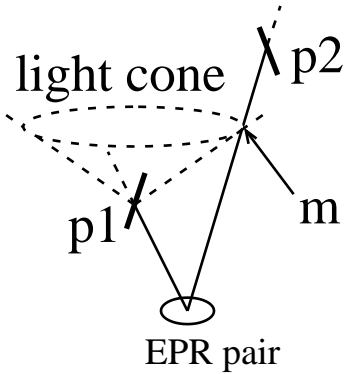}
\caption{Physical model of suggested in \cite{clinic} trial}
\label{eprt}
\end{center}
\end{figure}

Such kind of experiments are not performed (yet?) not only because of certain
difficulties, but also, because treatment of counterfactual data suggested
in \cite{clinic} is not very common in natural sciences.
Let us consider physical model correponding to first kind of trials
suggested in \cite{clinic} Fig.~\ref{eprt}.

Here should be considered not a simple case, then time gap between events
is very small (it would correspond to less accurate earlier tests
of quantum nonlocality, then space-like interval between events can be not
guaranteed), but as an experiment when between a moment ``{\bf m}'' (when it
is already possible to state, that fist photon was absorbed by
detector after first polarizer {\bf p1} and only one photon still exists) and
the second measurement {\bf p2} of polarization passed (in laboratory time)
even not few days or weeks (as it would be necessary for purposes discussed in
\cite{clinic}), but at least a few seconds.

Such kind of time-like quantum correlation still impracticable and not only
because of enormous difficulties, but also due to doubtful possibility of
clear interpretation of such experiments: on the one hand it could be
considered as strange example of quantum correlations between past and
future, between existing and not-existing particle, but there is more simple,
usual and mathematically equal model suggesting, that somewhere after
moment ``{\bf m},'' and before measurement {\bf p2} there is only one
particle in pure state described by polarization orthogonal to measured by
first polarizer {\bf p1}. And finally, there is (again mathematically
equivalent) absolutely weird model, that it is future measurement
of polarization {\bf p2} influences on first particle and causes it to be
in pure state with polarization orthogonal to result of
this measurement\footnote{On the other hand we may hear sometime, ``doctor
suggested the pills {\em because} it will help your,'' as if future help
causes some past event. So even such weird interpretations indirectly
present in our reasoning.}.

{\bf Problem 1.}
{\em Despite of some resemblance, (due to ``uncovering secrets of Nature'')
physics has certain difference with cryptography. It is often not enough to
suggest a model that reproduces behavior of some system. One proper key is
enough to uncover code, but here often exists many different explanations
and some of them may be absurd or completely irrelevant.}

\section{Breltmann's pills}

Dr.\ Breltmann\footnote{
Name used by Bell is slightly changed by some reasons.} \cite{Socks}
recently developed new kind of (quantum{\tiny $^{\circledR}$}) medicine for
curing {\em suggestibility}, but clinical tests produced absolutely negative
result --- all patients, who following his advice and used the pills
(very bitter) after testing demonstrated, that their level of
suggestibility is even higher,
than in group of people, who refused to use the tablets. But after such result
Dr.\ Breltmann realized, that the trials anyway demonstrated very strong
influence of his pills on human beings and suggested to use it for medical
treatment of {\em obstinacy}. He succeeds this time --- clinical trials
confirm, that 100\% of patients, who agreed to use the pills lately passed a
difficult test on completely absence of obstinancy.

The similar example was considered also in \cite{clinic}, but here it
is reproduced in such extreme form to recall corresponding Bell
consideration: in \cite{Socks} he discussed classical model of
correlation for two spin-1/2 particles, then probabilities of registration
are \cite{Socks}:
\begin{equation}
\left.\begin{array}{l}
P(up,up)=P(down,down) = \frac{|a - b|}{2\pi} \\
P(down,up)=P(up,down) =\frac12-\frac{|a - b|}{2\pi}
\end{array}\right\}
\label{saw}
\end{equation}
instead of
\begin{equation}
\left.\begin{array}{l}
P(up,up)=P(down,down) = \frac12{\left(\sin\frac{a - b}2\right)^2} \\
P(down,up)=P(up,down) =\frac12-\frac12{\left(\sin\frac{a - b}2\right)^2}
\end{array}\right\}
\label{sin}
\end{equation}
in quantum case and wrote next \cite{Socks}:
\begin{quotation}
\footnotesize
``Thus the {\em ad hoc} model does what is required of it (i.e., reproduces
quantum mechanical results) only at $(a-b)=0$, $(a-b)=\pi/2$ and $(a-b)=\pi$,
but not at intermediate angles.

[\ldots] Could we not be a little more clever, and device a model which
reproduces the quantum formulae completely? No. It cannot be done, so long
as action at a distance is excluded. This point was realized only subsequently.
Neither EPR nor their contemporary opponents were aware of it. Indeed the
discussion was for long entirely concentrated on the points $|a-b|=0$,
$\pi/2$ and $\pi$.''
\end{quotation}

It should be mentioned, that correlation functions
\begin{equation}
E(a,b)=P(up,up)+P(down,down)-P(down,up)-P(up,down)
\end{equation}
produced from classical \eq{saw} and quantum \eq{sin}
expressions have similar behavior with same minimal value
(anticorrelation) for $a-b=0$, maximal (correlation) for $|a-b|=\pi$,
an intermediate for $|a-b|=\pi/2$ (see picture in \cite{GisLoop1}).

{\bf Problem 2.}
{\em Even for simpler setup with particles and polarizers there are
certain problems with distinction between quantum and classical correlations.
For much more complicated situation with clinical research such
difference may be completely hidden (see below)}

\section{Identity of physical systems. Non-identity of patients.
What is the spin of patient?}

There is yet another property of classical Bell's model discussed above:
for any difference between two angles $\Delta=a-b$ there is some
value $\Delta_c$ that reproduces all probabilities on classical
model {\em for the fixed values of angles}. Three points mentioned
by Bell are special case, then $\Delta=\Delta_c$. So we may not
say, that quantum correlation are ``stronger'' than classical as it
sometime may be found in not very accurate popular presentations
(in quote above even suggested, that it was so for {\em any} papers
before Bell's works).

As it follows from \cite{clinic}, authors are aware about the problem,
but in the abstract may be found following sentence:
\begin{quote}
\footnotesize
``Quantum effects could fake an increase of the recovery rate
by about  $13\%$ although the drug would hurt as many patients as
it would help if everyone took it.''
\end{quote}
It may cause a wrong conclusion, that quantum correlation may produce
additional fake effect, stronger than analogue with classical ``Breltmann's
placebo pills'' discussed earlier. Really in \cite{clinic} $13\%$ is
value of specific expression for collective property of some group of people%
\footnote{Due to properties of Bell's classical model discussed
above, quantum corelations could not be ensured by arbitrary number of
trials with fixed parameters, like angles of polarizers.} with different
``angular parameters'' \cite{clinic} and so only loosely can be considered
as ``recovery rate'' (it is like situation, then instead of value of blood
pressure in clinic you got a puzzle like ``your pressure is 0.87 of mine
two years ago.'')

In \cite{clinic} also is not discussed an important consequences that may
destroy overall idea. In such a model \cite{clinic} angles $a, b$ are not
simply accessible or at least observable parameters like for experiments
with polarizers. On the other hand,
for any experiment with fixed angles and trial with one particular patient
always exists two angles for classical model, that completely reproduce
necessary correlations, i.e. arbitrary value between $-1$ and $+1$.

It is necessary to perform many experiments with particles and four
concrete fixed angles, to prove quantum nature of correlation. It is
not enough to use only one pair of angles. It is clear also, that
we should ensure that each polarizer may have only two given angles.
If it may be set in some unknown variety of different angles, then we
again may not hope to measure any true correlations.

So model suggested in \cite{clinic} could work only if all patients may be
characterized by the same four fixed angles. Seems the limitation is
enough to make the model not very realistic, but here is yet another
related point. There is some difference between \eq{sin} above and equations
used in \cite{clinic}. All angles in expressions differ in two times.
Really both expressions are valid. The difference is because in \cite{clinic}
used expressions for photon with spin-1, but in \cite{Socks} for
electron with spin-1/2.

So here is rather naive question: even if we accept that quantum model of
patient suggested in \cite{clinic} is true, {\em what is ``a spin of
patient''?} Or yet another question, let us suggest, that clinical trial
may be described by correlation with some complicated function, not
``cos'' in \cite{clinic} and the expression may not be explained
neither by classical nor but quantum hidden latent parameters. Could we use
the experimental data as a proof, that we find some new law of physics?
That is the difference between such idea and model in \cite{clinic}%
\footnote{It was rather pathetic question partially related with suggestion
in \cite{clinic} to use counterfactual statements. It is written, that it
is possible, if we have a model. But we may have also a model
of new physical law, that explains new kind of correlation function mentioned
above, or mathematical model of a nurse-esper, who directly recovered patients
without any medicine.}? The problem, that we have two different ways to
explain some phenomenon (recovering of patients): with direct causal
effects of medicine and without it, but with latent quantum phenomena.
It is not clear, how to distinguish those and why we should accept rather
vague quantum model developed in \cite{clinic}. So \ldots

{\bf Problem 3.}
{\em Elementary physical systems are equivalent. This property together
with possibility of ideal repetition with same setup of experimental
equipment is necessary condition for testing quantum correlations, but
it is doubtful, that the same is true for real patients and clinical trials.}

{\bf Problem 4.}
{\em It is even not such a big flaw to object using classical theory
for clinical trials (after all, ``classical'' is some synonym of
``known already to scientists living more than century ago''), but how does
it possible to use only simplest models of quantum theory (hardly
applicable even to two particles, see above) for very complex beings?}

\section{``$\bf C^*$ algebra of patient.''}

Some problems discussed above was concerned rather with difficulty of
observation quantum effects discussed in \cite{clinic}, but it may be
not must important problem. Authors of \cite{clinic} themselves were aware
about it and wrote in abstract:
\begin{quote}
\footnotesize
``We do not present any realistic model showing this effect, we only point
out that the physics of decision making could be relevant for the causal
interpretation of every-day life statistical data.''
\end{quote}
and in conclusion:
\begin{quotation}
\footnotesize
\ldots``Hence the violation of
the instrumental inequalities requires quantum coherence
that is stable for a long time (compared to time scales
of  decoherence  in technical quantum systems).
Therefore the violation seems to be even less likely.

However, the main purpose of this paper was to show that
some classical causal interpretations of  every-day life
statistical data could {\it in principle} fail
if latent quantum effects influence our behavior.''
\end{quotation}

But it was not mentioned rather simple principle:

{\bf Problem 5.}
{\em There are many absolutely different phenomena described by similar
expressions and so resemblance of mathematical formulae should not be
considered as proof of equality.}

After all it could be possible to comment some interesting
result discussed in \cite{clinic} as {\em it is
found, that models of linear algebra developed for description of quantum
mechanical systems are also convenient for description of decision making
in AI and statistics}, but not as an evidence of
reducibility of some psychological and physiological processes to quantum
entanglement of two particles somewhere inside of human brain (idea
about ``one-particle entanglement'' suggested in \cite{clinic} even
more mysterious).

\section{Affine separability. (ZHSL-BCJLPS theorems and
fake quantum correlations)}

May be pretensions about vague quantum models or irony about
spin of patient are not appropriate here? After all in \cite{clinic} is
discussed saturated and detailed $C^*$-algebraic model of patient
and it may include complicated issues of spins or quantum behavior
of multi-particle alive systems. But here there is yet another point.

As it was already mentioned in introduction, basic general question related
with \cite{clinic} is: how it is possible in principle to talk about
quantum correlations in clinical trials, if even in ``refined'' condition
of usual physical experiments, it is challenging task near limit of accuracy
of modern physical equipment? But if we have a complicated mathematical
model like in \cite{clinic}, how to check the misgivings about inadequacy?

Seems here may help a rather general result about ``affine separability''
discussed in \cite{afloop} and overlapped with earlier applications
of {\em ZHSL-BCJLPS theorem}\footnote{\.{Z}yczkowski, Horodecki, Sanpera,
Lewenstein \cite{SepVol} -- Braunstein, Caves, Jozsa, Linden, Popescu,
Schack \cite{SepNMR} theorem.}.

More detailed account may be found in original papers
\cite{SepVol,SepNMR,afloop}, but general idea rather simple and
shows, that in realistic conditions with weak signal due to quantum effect
and big background noise it is not possible to distinguish classical
and quantum correlations. Due to such problem in \cite{SepNMR} was
raised question about legitimacy to talk about quantum effect in real
NMR experiments. That can be said in such a case about clinical trials?

The trials considered in \cite{clinic}
are also relevant to error model necessary for wrong interpretation
used in \cite{afloop}.
Very close analogue of the models \cite{clinic} was considered in
\cite{casectrl}. It was discussed misclassification of
case-control studies\footnote{Studies discussed in \cite{clinic}
are called so sometime.} due to problems similar with discussed in
\cite{clinic}. The model developed in \cite{afloop} could not be mentioned
in \cite{casectrl}, written more earlier, but after developing of
this model and even before appearance of \cite{clinic} it becomes clear,
that such a model is relevant not only to physical research, but in
more wide area including situations discussed in \cite{casectrl}.

The model of error developed in \cite{afloop} also appropriate to general
algebraic framework used in \cite{clinic},
because let show, that using quite simple transformation it is possible
to produce ``classically correlated state'' from arbitrary quantum state
and it is relevant also for description of quantum states used in
\cite{clinic} (i.e., some linear functional on a $C^*$ algebra \cite{Murphy}).%
\footnote{It should be mentioned yet, that ZHSL-BCJLPS theorem formally still
proved only for finite-dimensional systems, but it is enough for consideration
of \cite{clinic}.}

The transformation corresponds to distortion of probabilities $p_k$ of
some outcomes described as
\begin{equation}
 p_k' = s p_k - b.
\label{aff}
\end{equation}
Here coefficients $s$ and $b$ should be connected by simple relation
to ensure unit probability for sum of all possible outcomes
(see \cite{afloop}).
The equation \eq{aff} may looks strange only at first sight, but it
has many realistic applications. For example, it corresponds to
linear approximation of a model with arbitrary nonlinear function
$
 p' = f(p)
$
or may be rewritten as
\begin{equation}
 p' = a\,p - b\,\bar p, \quad a = s-b, \quad \bar p \equiv 1 - p
\label{aff2}
\end{equation}
and the expression produces more clear interpretation of \eq{aff}.
It corresponds to erroneous reduction of probability estimation
due to events of ``complimentary''\footnote{In classical sense.} kind.

\begin{figure}[ht]
\begin{center}
\includegraphics{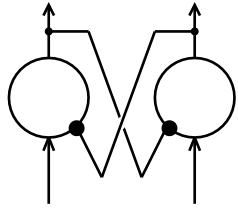}
\caption{Lateral inhibition}
\label{neu2}
\end{center}
\end{figure}

It should be mentioned also, that \eq{aff} is not necessary related with
some error. It may describe true change of probabilities of some processes
due to pure classical reason. Let us consider example with two neurons
considered in \cite{clinic}. There was suggested Bell-type correlations
between two neurons due to quantum effects. It is strange to use
ideas of nonlocality test with special setup ensuring of noninteraction
between objects to systems with possibility of active interdependency.
For neurons it is very actual, because ``saturation of probabilities''
discussed in relation with \eq{aff2} is very common process here \cite{neu}
and can be represented by simple scheme of {\em lateral inhibition}
Fig.~\ref{neu2}.

It may be familiar in physical society due to chapter about color
vision in first volume of {\em The Feynman lectures on physics} \cite{FLP1}.
Here mathematical problem with physically impossible {\em negative} value for
intensity of red color discussed in \cite{FLP1} has formal resemblance
with physically impossible negative probability discussed by Feynman in
his first lecture on quantum simulators \cite{FeySim}, and it is again
an example of mathematically similar and physically different models.
Excitations of neurons may be described by models of linear algebra
\cite{neu} and it is similar with quantum mechanics, but the excitation
is result of very difficult chemical processes with many molecules and
classical description.

\begin{figure}[ht]
\begin{center}
\includegraphics{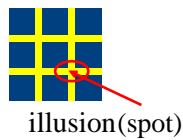}
\caption{Visual illusion}
\label{illus}
\end{center}
\end{figure}

For testing some of these laws are enough simple pictures like
Fig.~\ref{illus}.

{\bf Problem 6.}
{\em The model discussed in \cite{clinic} is illustrative example of conditions
for fake quantum correlations due to errors or wrong interpretation
of complicated experimental data recently discussed in \cite{afloop}.
So even appearance of regularity suggested by the models \cite{clinic}
could not confirm quantum correlation due to many unpredictable
sources of errors in real clinical trials, because in such conditions
criteria for distinction between classical and quantum correlations may not
be applied.}


\begin{thebibliography}{99}
\bibitem{clinic} D. Janzing and T. Beth,
\emph{e-print} \texttt{quant-ph/0208006} (2002).
\bibitem{Aspect} A. Aspect, J. Dalibard, and G. Roger,
\emph{Phys. Rev. Lett.} \textbf{49}, 1804 (1982).
\bibitem{GisLoop1}N. Gisin and B. Gisin,
\emph{Phys. Lett. A} \textbf{260}, 323 (1999);
\emph{e-print} \texttt{quant-ph/9905018}.
\bibitem{GisLoop2}N. Gisin and B. Gisin,
\emph{e-print} \texttt{quant-ph/0201077} (2002).
\bibitem{Socks}
J. S. Bell, ``Bertlmann's socks and the nature of reality,''
\emph{Journal de Physique} $\bf 42^3$ C2 41 (1981);
reprinted in: \emph{Speakable and unspeakable in quantum mechanics},
(Cambridge University Press, Cambridge, 1987).
\bibitem{SepVol}
K. \.{Z}yczkowski, P. Horodecki, A. Sanpera, and M. Lewenstein,
\emph{Phys. Rev. A} \textbf{58}, 883 (1998);
\texttt{quant-ph/9804024}.
\bibitem{SepNMR}
S. L. Braunstein, C. M. Caves, R. Jozsa, N. Linden, S. Popescu, and R. Schack,
\emph{Phys. Rev. Lett.} \textbf{83}, 1054 (1999);
\texttt{quant-ph/9811018}.
\bibitem{afloop}  A. Yu. Vlasov,
\emph{e-print} \texttt{quant-ph/0207013} (2002).
\bibitem{Murphy} G. J. Murphy, \emph{C$^*$-algebras and operator theory},
(Academic, Boston 1990).
\bibitem{casectrl} A. Shlyakhter, L. Mirny, A. Vlasov, and R. Wilson,
\emph{Human and Ecological Risk Assessment} $\bf 2^4$, 920 (1996).
\bibitem{neu} E. N. Sokolov, G. G. Vatkyavichus, \emph{The neurointelligence}
(Nauka, Moscow 1989) [and references therein].
\bibitem{FLP1} R. P. Feynman, R. B. Leighton, M. Sands, \emph{The Feynman
lectures on physics} {\bf Vol. 1}, (Wesley, Reading MA, 1963).
\bibitem{FeySim} R. P. Feynman,
\emph{Int. J. Theor. Phys.} \textbf{21}, 467 (1982).
\end{thebibliography}
\end{document}